\documentstyle[12pt,aaspp4]{article}  
\input epsf

\newcommand{\flux}{\mbox{ ergs cm}^{-2}\,\mbox{s}^{-1}}

\newcommand{\apc}{\mbox{ amu cm}^{-3}}  
\newcommand{\kms}{\mbox{ km s}^{-1}}  
  
\newcommand{\kpc}{\mbox{ kpc}}  
\newcommand{\vem}{\mbox{ ergs cm}^{-3}\,\mbox{s}^{-1}}
\newcommand{\La}{Ly$\alpha$}  
\newcommand{\Ha}{H$\alpha$}  
\newcommand{\NV}{N\,{\sc v} $\lambda\lambda1239,1243$}
\newcommand{\NII}{[N\,{\sc ii}] $\lambda\lambda6548,6584$} 
\newcommand{\OI}{[O\,{\sc i}] $\lambda\lambda6300,6364$} 
\newcommand{\SII}{[S\,{\sc ii}] $\lambda\lambda6717,6731$}

\begin{document}
\title{New HST Observations of High Velocity \La\ and \Ha\ 
in SNR 1987A}  
\author{Eli Michael\altaffilmark{1}, 
Richard McCray\altaffilmark{1}, 
C. S. J. Pun\altaffilmark{2},   
Kazimierz Borkowski\altaffilmark{3}, 
Peter Garnavich\altaffilmark{4}, 
Peter Challis\altaffilmark{4},    
Robert P. Kirshner\altaffilmark{4}, 
Roger Chevalier\altaffilmark{5},
Alexei Fillippenko\altaffilmark{6},
Claes Fransson\altaffilmark{7},
Nino Panagia\altaffilmark{8},
Mark Phillips\altaffilmark{9},
Brian Schmidt\altaffilmark{10},
Nicholas Suntzeff\altaffilmark{9},
and J. Craig Wheeler\altaffilmark{11}}

\altaffiltext{1}{JILA, University of Colorado, Boulder, CO 80309-0440;
michaele@colorado.edu; dick@jila.colorado.edu}
\altaffiltext{2}{Laboratory for Astronomy and Space Physics, Code 681,
NASA - GSFC, Greenbelt, MD 20771; pun@congee.gsfc.nasa.gov}
\altaffiltext{3}{Dept. of Physics, North Carolina State University,
Raleigh, NC 27695; kazik@mozart.physics.ncsu.edu}
\altaffiltext{4}{Harvard-Smithsonian Center for Astrophysics, 60
Garden St, Cambridge, MA 02138; (pgarnavich,pchallis,kirshner)@cfa.harvard.edu}
\altaffiltext{5}{Dept. of Astronomy, University of Virginia, P.O. Box
3818, Charlottesville, VA 22903-0818}
\altaffiltext{6}{Dept. of Astronomy, University of California,
Berkeley, CA 94720}
\altaffiltext{7}{Stockholm Observatory, S-133 36 Saltsj$\ddot{\rm{o}}$baden, 
Sweden}
\altaffiltext{8}{STScI, 3700 San Martin Drive, Baltimore, MD 21218}
\altaffiltext{9}{CTIO, NOAO, Casilla 603, La Serena, Chile}
\altaffiltext{10}{Mount Stromlo and Siding Spring Observatory, Private
Bag, Weston Creek P.O., Australia}
\altaffiltext{11}{Dept. of Astronomy and McDonald Observatory,
University of Texas, Austin, TX 78712}

\begin{abstract} 
We describe and model high velocity ($\approx 15,000 \kms$) \La\ and
\Ha\ emission from supernova remnant 1987A seen in September and
October 1997 with the Space Telescope Imaging Spectrograph.  Part of
this emission comes from a reverse shock located at $\approx 75\%$ of
the radius of the inner boundary of the inner circumstellar ring and
confined within $\pm 30^\circ$ of the equatorial plane.  Departure
from axisymmetry in the \La\ and \Ha\ emission correlates with that
seen in nonthermal radio emission and reveals an asymmetry in the
circumstellar gas distribution. We also see diffuse high velocity
\La\ emission from supernova debris inside the reverse shock that may
be due to excitation by nonthermal particles accelerated by the shock.
\end{abstract}


\section{Introduction}

On May 24, 1997 (day 3743 after core collapse), Sonneborn {\it et al.}
(1998) observed supernova remnant (SNR) 1987A with the Space Telescope
Imaging Spectrograph (STIS). They discovered very broad ($\approx
20,000 \kms$) \La\ emission in the ultraviolet spectrum.  Michael {\it
et al.} (1998, hereafter M98) interpreted this emission as the result
of excitation of neutral hydrogen atoms in the supernova debris
crossing a reverse shock.  The observed \La\ flux was close to the
value predicted by Borkowski {\it et al.}  (1997, hereafter B97) for
such a model normalized to account for the soft X-ray flux observed by
ROSAT.  M98 also pointed out that the broad \La\ emission should be
accompanied by \Ha\ emission that should be observable both from the
ground and with STIS.

The observations reported by Sonneborn {\it et al.} (1998) were taken
with a $2\arcsec\ \times 2\arcsec$ aperture that included the entire
inner circumstellar ring of SNR 1987A.  With 24 \AA\ arcsec$^{-1}$
dispersion, the resulting \La\ image seen on the STIS detector is a
convolution of the spatial ($\approx 1\arcsec$) and velocity ($\approx
3\arcsec$) structure of the emitting region.  M98 found that they
could fit the observed image with a model in which the shock surface
was a prolate ellipsoid slightly elongated in the direction normal to
the ring plane, but they found evidence for an additional source of
diffuse emission inside the shock surface.  This result was
surprising, because the model in B97, following Chevalier \& Dwarkadas
(1995), implies that most of the emission should come from a ring
shaped surface inside the inner circumstellar ring.

Here we report a new set of STIS observations of the fast shocks in
SNR 1987A, this time taken with slits covering only part of the
emitting region and including \Ha\ as well as \La. These new
observations enable us to distinguish the dynamics and morphology of
the fast shocks more clearly.  We now see that the shock emission is
indeed concentrated in the equatorial plane, contrary to our
interpretation of the earlier STIS observations.  We also see a
diffuse component of emission that may indicate excitation and
ionization by nonthermal particles.  We have not yet detected an
unambiguous signal from high velocity \NV\ emission.

The same set of observations also provides new information on the
spectra of the supernova ejecta and the circumstellar ring, including
a rapidly brightening ``hot spot'' that is apparently caused by the
first encounter of the supernova blast wave with the ring.  We will
interpret those data in subsequent papers.

\section{Observations}

The Supernova Intensive Study (SINS) collaboration obtained STIS
spectra of SNR 1987A in ultraviolet and optical wavelengths in
September and October 1997 respectively. The ultraviolet spectrum,
which covers the wavelength range 1130 -- 1720 \AA, was taken on JD
2,450,719.1, or 3869.3 days since explosion, with the G140L low
dispersion grating ($\Delta v \approx 300 \kms$ at \La).  The
resolution along the spatial direction is $\approx 0\farcs06$.  Five
spectral images totaling 11200 sec of exposure were combined.  The
optical spectrum, which covers the wavelength range 5260 -- 10260 \AA,
was taken on JD 2,450,727.8, or 3878.0 days since explosion, with the
G750L low resolution grating ($\Delta v \approx 450 \kms$ at \Ha). The
spatial resolution is $\approx 0\farcs1$.  Two spectral images
totaling 4800 sec of exposure were obtained.  The two images were
dithered along the orientation of the slit so that both cosmic ray
hits and hot pixels on the STIS/CCD were removed simultaneously in the
final combined data.

Figure \ref{fig1}a illustrates the slit positions on the inner circumstellar
ring for the two observations. A $0\farcs5$ slit was used for \La\ and
a $0\farcs2$ slit was used for \Ha.  Both slits were offset $0\farcs1$
to the SE of the centroid of the ring, with position angles of
$-147^\circ$ and $-139^\circ$ for the \La\ and \Ha\ observations,
respectively.

\bigskip 
\centerline{\epsfxsize=0.5\hsize{\epsfbox{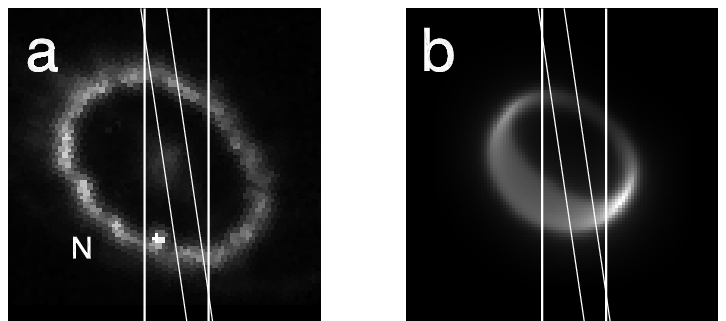}}}
\figcaption{(a) The position of the \La\ (wide) and \Ha\ (narrow)
slits on the inner ring, $\bf{N}$ denotes the northern side of the
remnant.  (b) position of the slits on the reverse shock surface which
lies at $\approx 70\%$ of the distance to the inner ring.\label{fig1}}
\bigskip 

Figure \ref{fig2}a shows the ultraviolet spectrum in the vicinity of \La.  It
displays the following components: (1) bright stationary geocoronal
\La\ emission filling the slit; (2) a nearly horizontal streak of
intense blue-shifted \La\ emission below the centerline but inside the
lower (NE) side of the ring extending to $-15,000 \kms$; (3) a fainter
horizontal streak of red-shifted emission above the centerline but
inside the upper (SW) side of the ring extending to $12,000 \kms$; (4)
a vertical band at $4000 - 8000 \kms$ due to diffuse, \NV\ emission
from nearly stationary N\,{\sc v} ions in the vicinity of the ring;
and (5) a broad component of diffuse emission extending from $-
15,000$ to $20,000 \kms$.

Figure \ref{fig2}c shows the optical spectrum in the vicinity of \Ha.  It
displays the following components: (1) pairs of bright spots above and
below the centerline due to emission by the nearly stationary inner
ring at \OI, \NII, \Ha, and \SII; (2) broad horizontal bands along the
centerline due to emission in these same lines from atoms in the inner
supernova debris excited by radioactivity; and (3) a nearly horizontal
streak of blue-shifted \Ha\ emission below the centerline but inside
the circumstellar ring extending to $-12,000 \kms$.

\bigskip
\centerline{\epsfxsize=0.5\hsize{\epsfbox{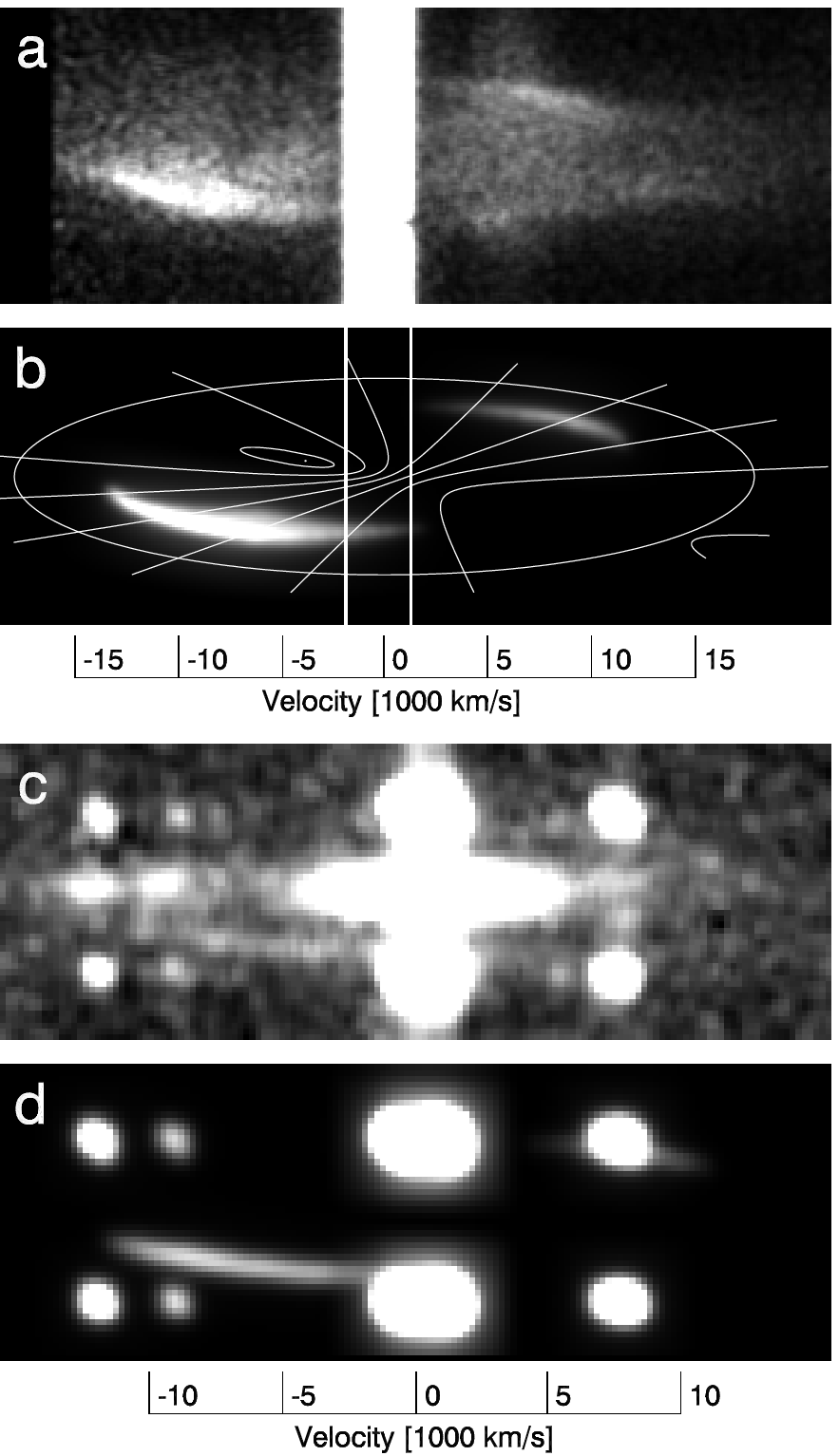}}} 
\figcaption{ (a) STIS G140L spectrum of SNR 1987A in the 
vicinity of \La. (b) Simulated STIS image of \La\ emission 
from the reverse shock surface. (c) STIS G750L spectrum of 
SNR 1987A centered on \Ha. (d) Simulated STIS image of line 
emission from the stationary ring and \Ha\ emission from 
the reverse shock surface. The velocity scales are line of 
sight velocity measured from LMC rest frame.\label{fig2}} 
\bigskip

\section{Reverse Shock Emission}

According to B97 and M98, we expect to see \La\ and \Ha\ emission from
H\,{\sc i} atoms in the supernova ejecta that are excited by electron
and ion impacts as they cross a reverse shock.  (We do not expect to
see \La\ and \Ha\ emission from the blast wave because it is entering
a region where the hydrogen atoms are already nearly fully ionized.)
Since the H\,{\sc i} atoms are freely streaming, with radial velocity
${\bf v} = {\bf r}/t$, before they cross the shock surface, the
Doppler shifts of the \La\ and \Ha\ are determined by the geometry of
the surface. M98 inferred from the earlier STIS observations that the
reverse shock surface was a slightly prolate ellipsoid.  For such a
model, they predicted that a STIS observation of \La\ through a
$0\farcs5$ slit would appear as an elliptical image (their Fig. 1d).

The actual STIS images of \La\ and \Ha\ (Figs. \ref{fig2}a,c) do not agree well
with the prolate ellipsoid model for the reverse shock. Instead, they
agree better with the model proposed by B97, in which the reverse
shock emission is confined primarily to an equatorial band.  We find
also that the position and shape of the reverse shock surface vary
from the far to the near side of the remnant, and probably on smaller
scales as well.

We can fit both the blue-shifted \La\ and \Ha\ observations with a
model in which the shock on the near side has a radius which ranges
from $0.73 R_{ring}$ on the left side of the $0\farcs5$ slit to $0.70
R_{ring}$ on the right side of the slit and extends in latitude from
$30^\circ$ to $- 20^\circ$.

To fit the red-shifted \La\ observations, the shock surface on the far
side must have a radius $0.76 R_{ring}$ and extend in latitude from
$10^\circ$ to $-5^\circ$.  The red-shifted \Ha\ emission was too
faint to measure.

Figure \ref{fig1}b illustrates the reverse shock surface and its
orientation in the slits for the \La\ and \Ha\ observations.  The
surface outside of the slits is not constrained by the present data
and is shown only as an extrapolation.  Figures \ref{fig2}b,d show the
resulting model \La\ and \Ha\ STIS images, respectively. Lines of
constant latitude are shown on the simulated \La\ image assuming that
all the emission comes from the center of the slit (see \S 4).
Comparison of the model \La\ and \Ha\ images demonstrates the
advantage of a narrow slit in mapping the shock surface.

When we subtract the simulated images from the actual observations,
the streaks vanish, leaving residual emission in the \La\ observation,
which we discuss below in \S 4.  The measured flux from the \La\ shock
emission is $(8.6 \pm 2.6) \times 10^{-15} \flux$ on the blue side and
$(1.8 \pm 0.8) \times 10^{-15} \flux$ on the red side.  The measured
flux from the \Ha\ shock emission is $(6.0 \pm 1.8) \times 10^{-16}
\flux$ on the blue side and $< 4 \times 10^{-16} \flux$ on the red
side.

The blue-shifted \La\ flux from the shock is greater than the
red-shifted flux by a factor of $\approx 4.8$. If we assume
that the outer SN debris is spherically symmetric with
a power law density profile given by $\rho(r,t) \propto t^{-3}
(r/t)^{-9}$ (Eastman \&\ Kirshner 1989), we can attribute this
asymmetry to: the difference in shock radius (which contributes a
factor $\approx 1.7$); the fact that the reverse shock has a greater
surface area on the near side (a factor $\approx 2.1$), and the
$\approx 0.3$ year difference in light-travel time from the near side
of the shock surface to the far side (a factor $\approx 1.3$).

This asymmetry should not come as a great surprise.  Gaensler {\it et
al.} (1997) found that the nonthermal radio emission seen inside the
inner circumstellar ring was much brighter on the northeast side than
on the southwest side. The asymmetry in the reverse shock surface is
not caused by an asymmetry in the supernova debris because then the
red-shifted \La\ emission would be brighter due to the higher mass
flux (ram pressure) required to drive the shock to a greater radius.
The asymmetry therefore represents an asymmetry in the circumstellar
gas distribution which the ejecta is encountering (the circumstellar
gas on the far side must be more tenuous than the near side).

M98 calculated that, for the conditions behind the reverse shock,
approximately 1 \La\ and 0.2 \Ha\ photons will be emitted per H\,{\sc
i} atom entering the shock.  We can infer this emissivity ratio from
the observations and our model, which allows us to correct for the
fact that the \La\ observation is taken through a wider slit than the
\Ha\ observation. Correcting for interstellar extinction factors of
0.15 at \La\ and 0.63 at \Ha\ (Fitzpatrick {\it et al.}  1990) we
find, on the blue-shifted side, where the \Ha\ streak is bright enough
to measure, this ratio to be $4.6 \pm 2$, consistent with our
expectation.

Since we know the number of \La\ photons emitted per H\,{\sc i} atom
crossing the shock surface, we can infer the normalization [$A$, the
density at $(r/t) = 10,000 \kms$ and $t = 10$ years] of the SN
envelope's density profile from our model for the shock
surface. Estimating that the reverse shock is propagating outward with
velocity $V_{rs} = 4,000 \kms$ (B97) and assuming a helium number
density $n_{\rm{He}} = 0.2n_{\rm{H}}$, we find $A = 170 \pm 60 \apc$,
which may be compared with Shigeyama \& Nomoto's (1990) model 14E1 ($A
= 120 \apc$) and Woosley's (1988) model 10H ($A = 360 \apc$).  If we
had taken the value of $V_{rs}$ inferred from radio observations of the
remnant (Gaensler {\it et al.} 1997), $A$ would be decreased by
$\approx 10\%$.

\section{Diffuse Emission}

When we subtract the model images from the actual images, a diffuse
component of high velocity \La\ emission remains.  This component was
evident in the earlier slitless observation by Sonneborn {\it et al.}
(1998), and led M98 to interpret those data with a model in which the
reverse shock surface had the shape of a slightly prolate ellipsoid.
Now, with the present observations, we can see clearly that the \La\
and \Ha\ emission from the reverse shock is confined to the equatorial
plane, and that the residual \La\ emission comes from a volume
interior to that surface. A corresponding diffuse component of \Ha\
presumably exists, but it is too faint to be seen with the present
exposure.

To analyze the distribution of the diffuse \La\ emission, we assume
that the emitting gas is everywhere in free expansion.  If so, and if
the slit were very narrow, there would be a unique mapping from the
STIS image to depth within the supernova debris.  From such a mapping,
we could reconstruct a two-dimensional image of the supernova debris
intercepted by the slit.  Figures \ref{fig3}a-c are just such images,
where the coordinates are radius and latitude measured from the
equatorial plane of the inner circumstellar ring.  The curves of
constant latitude are not radial lines because the slit is off-center.

\bigskip
\centerline{\epsfxsize=\hsize{\epsfbox{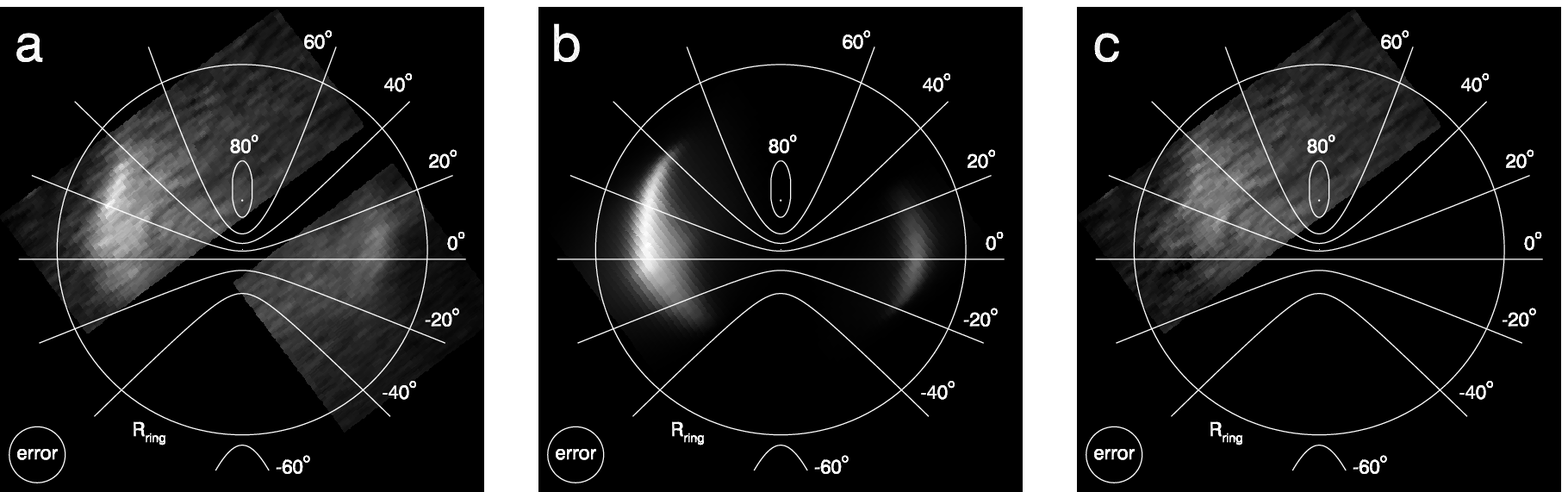}}}
\figcaption{Remapping of \La\ emission where appropriate, see 
text (\S 4). (a) observed emission, (b) simulated reverse shock 
emission, and (c) interior blue-shifted emission (panel (b) subtracted 
from panel (a)).\label{fig3}}
\bigskip

Figure \ref{fig3}a represents the \La\ data of Figure \ref{fig2}a in
such a coordinate system, in which we assume that the slit is a line
centered in the actual slit.  Since the actual slit has width
$0\farcs5$, the mapping is not unique; this introduces a
``point-spread function'' for the mapping represented approximately by
the error circles in Figures \ref{fig3}a-c.

Since we cannot distinguish the contribution of red-shifted \La\ from
that of \NV\ with the present data, we cannot interpret the signal on
the lower right hand side of Figure \ref{fig3}a.

Figure \ref{fig3}b represents the model for \La\ emission from the
reverse shock surface corresponding to Figure \ref{fig1}b, including
the broadening due to the $0\farcs5$ slit.  In this coordinate system,
we can see clearly that the blue-shifted \La\ emission comes from a
reverse shock surface that lies primarily above the equatorial plane.

Figure \ref{fig3}c shows the residual flux on the blue side of \La\
after subtracting the contribution from the reverse shock from the
data.  This diffuse \La\ emission comes from supernova debris inside
the reverse shock surface and appears to have a roughly uniform radial
distribution.  Like the \La\ emission from the reverse shock surface,
the diffuse \La\ emission is strongest within latitudes $0^\circ$ to
$40^\circ$; but we see significant diffuse emission from $-20^\circ$
to $90^\circ$.

The flux of diffuse blue-shifted \La\ seen through the $0\farcs5$ slit
is $(5.7 \pm .3) \times 10^{-14} \flux$, or $\approx 7$ times the flux
of blue-shifted \La\ from the reverse shock. Each pixel in the
spectrum observes a volume element in the debris so we are able to
measure the volume emissivity of the diffuse interior
emission. Correcting for interstellar extinction and assuming a
distance of $50 \kpc$ to the remnant we find this emissivity to be
$\approx 4 \times 10^{-19} \vem$.

What causes this diffuse emission?  Since it is concentrated in the
same latitude range as the reverse shock emission, one is tempted to
look at the reverse shock itself as the source of the excitation.  We
can rule out one obvious candidate --- soft X-rays emitted by the
shock --- because the soft X-ray luminosity observed by ROSAT
(Hasinger et al. 1996) is roughly an order of magnitude too faint to
account for the diffuse \La.  Likewise, the hot spot is still too
faint to cause significant diffuse \La\ emission. We suspect that the
diffuse \La\ emission is caused by nonthermal particles that are
accelerated at the reverse shock and diffuse into the freely expanding
supernova envelope.  We know that such particles are present from
observations of nonthermal radio emission (Gaensler {\it et al.}
1997).  The reverse shock may be a fairly efficient source of
nonthermal particles (Jones \&\ Ellison 1991).  Moreover, those
nonthermal particles that diffuse into the supernova envelope and
deposit their energy may produce \La\ photons with efficiency
approaching $\approx 30\%$.  

\section{Future Investigations}

Further observations with STIS will give us a much clearer picture of
the complex dynamics and kinetics of this extraordinary event.  It
would be valuable to observe the system again in the same slit
configurations to measure the rate of brightening in \La\ and \Ha\
emission from the shock surface.  We should map the high velocity
emission from the entire system, especially near the east side where
the radio lobes are brightest.

As M98 have emphasized, observations of the line profile of high
velocity \NV\ emission will provide a unique probe of the kinetics of
a collisionless shock.  We believe that the emission seen on the red
side of \La\ contains a significant contribution from \NV, but we
cannot distinguish this emission from \La\ emission in the present
observations.  To do that, we must observe the high velocity emission
from the same part of SNR 1987A with STIS in both \La\ and \Ha.  The
exposure should be deep enough to see the diffuse emission clearly in
\Ha\ as well as \La, and the slit should be narrow in order to
separate the spatial structure from the velocity structure and to
isolate the \NV\ emission from stationary gas.  Then, since the ratio
of \La/\Ha\ should be constant, we would be able to subtract the \Ha\
image from the \La\ image, leaving a remainder that is due to \NV.

\acknowledgements This research was supported by NASA through grant
No. GO-2563.01-87A from the Space Telescope Science Institute, which
is operated by the Association of Universities for Research in
Astronomy, Inc., under NASA contract NAS 5-26555, and by NASA grants
NAG 5-3313 to the University of Colorado and 5-2844 to North Carolina
State University. C. S. J. P. acknowledges funding by the STIS IDT
through the National Optical Astronomy Observatories, and by the
Goddard Space Flight Center.


\begin{thebibliography}{}
\bibitem [Borkowski, Blondin, \&  McCray, 1997] {B97} 
Borkowski, K., Blondin, J., \& McCray, R.  1997, ApJ, 476, 
L31 (B97)
\bibitem [Chevalier \& Dwarkadas 1995] {CD95} Chevalier, R. 
A., \& Dwarkadas, V. V. 1995, ApJ, 452, L45 
\bibitem [Eastman \& Kirshner 1989] {EK89} Eastman, R. G., 
\& Kirshner, R. P. 1989, ApJ, 347, 771
\bibitem [Fitzpatrick \& Walborn 1990] {FW90} Fitzpatrick, 
E. L., \& Walborn, N. R. 1990, AJ, 99, 1483
\bibitem [Gaensler et al 1997] {G97}  Gaensler, B.  M., 
Manchester, R. N., Stavely-Smith, L., Tzioumis, A.  K., 
Reynolds, J. E., \& Kesteven, M. J. 1997, ApJ, 479, 845
\bibitem [Hasinger et al 1996] {H96} Hasinger, G., 
Aschenbach, B., \& Tr$\ddot{\rm{u}}$mper, J. 1996, A\&A, 
312, L9
\bibitem [Jones \& Ellison 1991] {JE91} Jones, F. C., \&\  
Ellison, D. C., Space Sci. Rev., 58, 259
\bibitem [Michael et al 1998] {M98} Michael, E., McCray, 
R., Borkowski, K. J., Pun, C. S. J., Sonneborn, G. 1998, 
ApJ, 492, L143 (M98)
\bibitem [Shigeyama \& Nomoto 1990] {SN90} Shigeyama, T., 
\& Nomoto, K. 1990 ,ApJ, 360, 242
\bibitem [Sonneborn et al 1998] {Sonn98} Sonneborn, G., {\it et al.}
1998, ApJ, 492, L139
\bibitem [Woosley 1988] {W88} Woosley, S. E. 1988, ApJ, 
330, 218 
\end{thebibliography}
\end{document}